\documentclass{ws-procs9x6}
\usepackage{amsmath,bbm}
\usepackage{amssymb}

\newcommand{\Se}{\Sigma_{\epsilon}}
\newcommand{\Ge}{\Gamma_{\epsilon}}
\newcommand{\G}{\Gamma}

\newcommand{\va}{\scriptscriptstyle}

\newcommand{\R}{\mathbb{R}}

\newcommand{\Hp}{{H}_{phys}}
\newcommand{\Hk}{{H}_{kin}}
\newcommand{\He}{{H}_{\epsilon}}

\newcommand{\PP}{{ P}}
\newcommand{\So}{{\hat {S}}}

\newcommand{\be}{\nopagebreak[3]\begin{equation}}
\newcommand{\ee}{\end{equation}}
\newcommand{\ba}{\nopagebreak[3]\begin{eqnarray}}
\newcommand{\ea}{\end{eqnarray}}

\begin{document}

%%%%%%%%%%%%%%%%%%%%%%%%%%%%%%%%%%%%%%%%%%%%%%%%%%%%%%%%%%%%%%
% title, author(s) and address(es) put here:                 %
%%%%%%%%%%%%%%%%%%%%%%%%%%%%%%%%%%%%%%%%%%%%%%%%%%%%%%%%%%%%%%

\title{Dynamics of loop quantum gravity and spin foam models in three dimensions}

\author{Karim Noui and Alejandro Perez}
\address{Center for Gravitational Physics and Geometry,
Pennsylvania State University, University Park, PA 16802, USA}


\maketitle

\abstracts{ We present a rigorous regularization of Rovellis's
generalized projection operator in canonical 2+1 gravity. This work
establishes a clear-cut connection between loop quantum gravity and the spin 
foam approach in this simplified setting. The point of view adopted here 
provides a new perspective to tackle the problem of dynamics in the physically 
relevant 3+1 case.}

%%%%%%%%%%%%%%%%%%%%%%%%%%%%%%%%%%%%%%%%%%%%%%%%%%%%%%%%%%%%%
% The main text of your paper                               %
%%%%%%%%%%%%%%%%%%%%%%%%%%%%%%%%%%%%%%%%%%%%%%%%%%%%%%%%%%%%%

\section{Introduction}

The goal of the {\em spin foam} approach \cite{a18} is to
construct a mathematically well defined notion of path integral for
non-perturbative quantum GR as a devise for computing the `{\em
dynamics}' of the theory. By `{\em dynamics}' here we mean the
characterization of the kernel of the quantum constraints given 
by the representation of the classical constraints in connection variables, schematically, $G_i = D_a
E^a_i$, $V_b = E^a_i F^i_{ab}$, and $S = E^a E^b
F_{ab}(A)$ \footnote{The field $A$ is an $SU(2)$ connection and $E$ is its canonical momentum represented by the 
sensitized triad both defined on a 3-manifold $\Sigma$ defining space. Space-time is assume to be of the topology $\Sigma\times \R$.}.

Spin foam models have been studied as an attempt to give an
explicit construction of the generalized projection operator $\PP$
from the kinematical Hilbert space $\Hk$, where the constraints above are defined as operators,
into their kernel $\Hp$. A tentative regularization of the formal projector into the kernel
of $S(x)$ and $V^a(x)$
\begin{equation}\label{P} P=``\prod_{x \in \ \Sigma} \delta(\hat V^a(x))\delta(\So
(x))"=\int D[N_{\mu}] \ {\rm exp}(i\int \limits_{\Sigma} N_{0}
\hat { S}+N_{a} \hat { V}^{a})\end{equation} was
presented in \cite{c2}\footnote{One can also define the notion of path integral for
gravity as a lattice discretization of the formal path integral
for GR in first order variables
\begin{equation}\nonumber
P=\int \ D[e]\  D[A]\ \mu[A,e] {\rm exp}\left[ i S_{\va GR}(e,A)
\right]
\end{equation}
\vskip-.1cm \noindent where the formal measure $\mu[A,e]$ must be
determined by the Hamiltonian analysis of the theory.}.
Given two {\em spin network} states $s,s^{\prime}$ the
physical scalar product $<s,s^{\prime}>_{phys}:=<Ps,s^{\prime}>$ can be
formally defined by \vskip-.3cm
\begin{equation} \label{vani}<s,s^{\prime}>_{phys}= \left<Ps,
s^{\prime}\right>=\int {D}[N] \sum \limits^{\infty}_{n=0}
\frac{i^{n}}{n!}<\left[\int \limits_{\Sigma} N(x) \hat {
S}(x)\right]^n \ s, s^{\prime}>,
\end{equation}
\vskip-.1cm \noindent where the exponential in (\ref{P}) has been
expanded in powers. From early on, it was realized that smooth
loop states are naturally annihilated (independently of any
regularization ambiguity) by $\hat { S}$ \cite{jac,c8}.
Consequently, $\hat S$ acts only on {\em spin network} nodes.
Generically, it does so by creating new links and nodes modifying
in this way the underlying graph of the {\em spin network} states.
The action of $\So$ can be visualized as an `interaction vertex'
in the time evolution of the node (see diagram on the right of Figure \ref{spinn}). 
Therefore, each term in the sum (\ref{vani}) represents a series of transitions--given by
the local action of $\hat { S}$ at {\em spin network}
nodes--through different {\em spin network} states interpolating
the boundary states $s$ and $s^{\prime}$ respectively. They can in
fact be expressed as a sum over `histories' of {\em spin network}
that can be pictured as a system of `colored' branching surfaces described
by a 2-complex labeled by spins. Every such history is called a {\em spin foam}.

Before even considering the issue of convergence of (\ref{vani}),
the problem with this definition is evident: every single term in
the sum is a divergent integral! Therefore, this way of presenting
{\em spin foams} has to be considered as formal until a well
defined regularization of (\ref{P}) is provided. Possible
regularization schemes are discussed in \cite{c2} although they
have not been implemented in concrete examples.

%The underlying discreteness discovered in loop quantum gravity is
%crucial: in {\em spin foam} models the functional integral for
%gravity is replaced by a sum over amplitudes of combinatorial
%objects given by foam-like configurations ({\em spin foams}). The
%precise definition was first introduced by Baez in \cite{baez5}. A
%{\em spin foam} represents a possible history of the gravitational
%field and can be interpreted as a set of transitions through
%different quantum states of space. Boundary data in the path
%integral are given by polymer-like excitations ({\em spin network}
%states) representing $3$-geometry states in loop quantum gravity.

Although many spin foam models for 4-dimensional gravity have 
been proposed\cite{a18}, the rigorous connection with the well developed
canonical theory of loop quantum gravity is still under investigation.
Here we show that a clear-cut connection between the loop approach and spin 
foams can be established in three dimensions.

Pure gravity in three dimensions is a well studied example of
integrable system that can be rigorously quantized. 
The reduced phase space of the theory is finite dimensional  and there 
exist different quantization schemes that make use of this property.

From our perspective $3$-dimensional gravity is taken as an toy
model for the application of quantization techniques that are
expected to be useful in four dimensions. In this sense we
want to quantize the theory according to Dirac prescription which
implies having to deal with the infinitely many degrees of
freedom of a field theory at the kinematical level, i.e., we want
to quantize first and then reduce at the quantum level. This is
precisely the avenue that is explored by loop quantum gravity in
four dimensions where the reduced phase space approach seems hopeless.

\section{Canonical three dimensional gravity}

Assuming the topology of space time to be of the form $M=\Sigma\times \R$, where
$\Sigma$ is an arbitrary Riemann surface,
the phase space of 3-dimensional gravity is parametrized the
2-dimensional connection $A_a^{i}$ and the triad field
$E^b_j$ where $a=1,2$ are $\Sigma$-coordinate 
indices and $i,j=1,2,3$ are $su(2)$ indices. The
symplectic structure is defined by $\{A_a^{i}(x), E^b_j(y)\}=\delta_a^{\, b} \delta^{i}_{\, j} \delta(x,y)$.
Local symmetries of the theory are generated by the Gauss and
curvature constraints, $D_b E^b_j=0$ and $F_{ab}(A)=0$, respectively.

\begin{figure}[h]\!\!\!\!\!\!
\centerline{\hspace{0.5cm} \(\begin{array}{c}
\includegraphics[height=3cm]{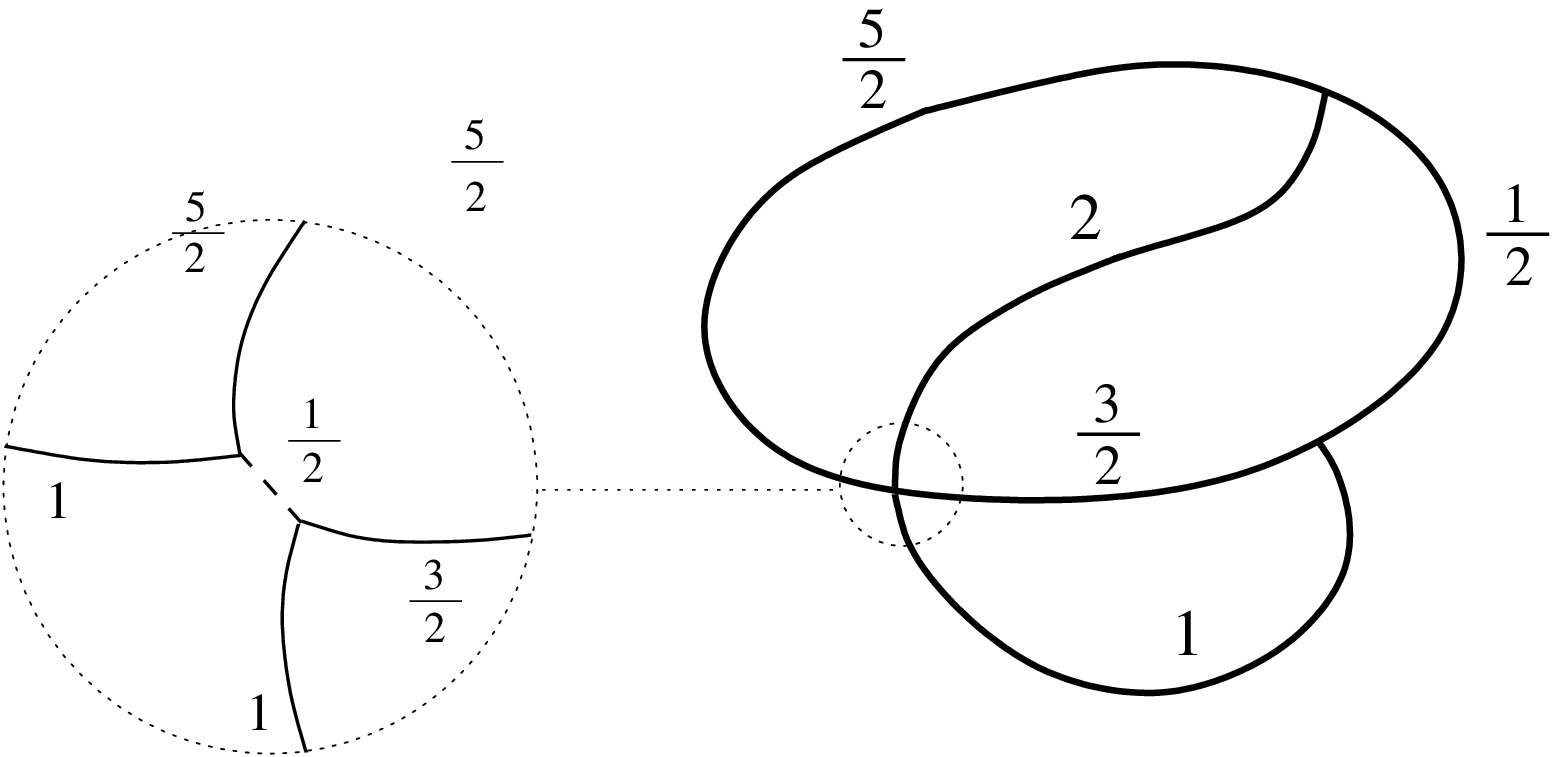}
\end{array}\ \  \ \ \ \
\begin{array}{c}
\includegraphics[height=3cm]{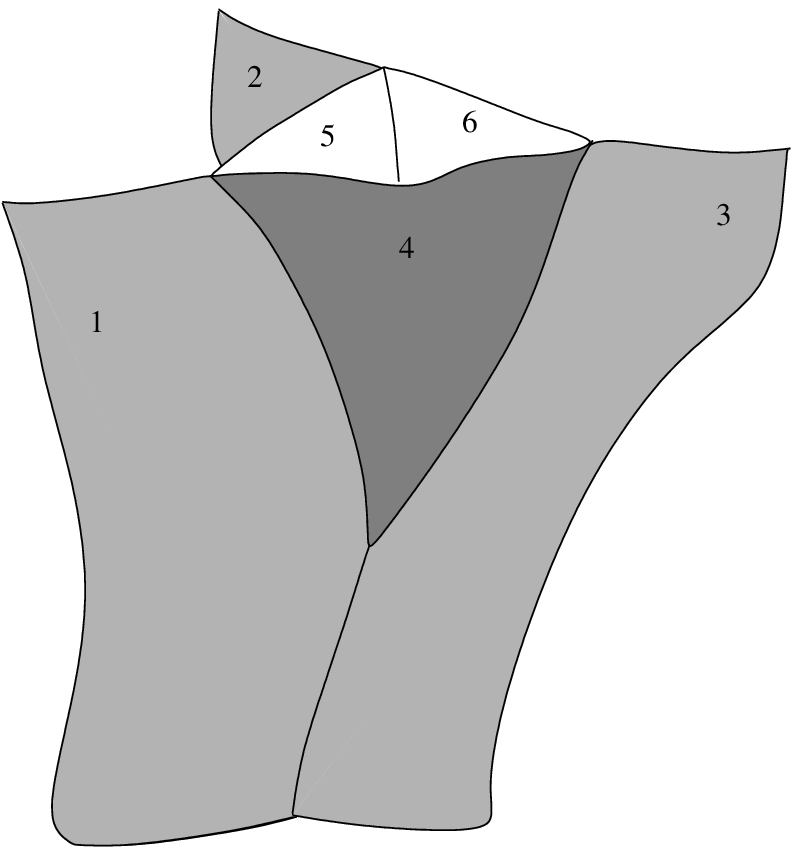}
\end{array} \) }
\caption{Spin-network state on the left. On the right `time' evolution
of a {\em spin-network node} into a {\em spin foam} vertex.} \label{spinn}
\end{figure}

As in 4d, the kinematical Hilbert space, $\Hk$, is given by a certain set of
$SU(2)$ gauge invariant functionals of the connection $\Psi[A]$
which are square integrable with respect to a natural
diffeomorphism invariant measure, the Ashtekar-Lewandowski (AL) measure
\cite{ash3,ash4}. As in standard gauge theories, the basic gauge
invariant functional of $A$ is given by the Wilson loop; namely,
the trace of the holonomy of $A$ around a close loop on the
fundamental representation of $SU(2)$. Any state in $\Hk$ can be
expressed in terms of linear combinations of products of Wilson
loops. Natural orthonormal basis of $\Hk$ can be constructed in
terms of eigenstates of geometric operators of quantum geometry
representing area and length, given by the so-called {\em spin
network} states \cite{reis8,c4,baez10}. {\em Spin network states} $\Psi_{\va \gamma,
\{j_{\ell}\},\{\iota_{n}\}}[A]$ are defined by a graph
$\gamma$ in $\Sigma$, a collection of spins $\{j_{\ell}\}$
associated with links $\ell\in \gamma$, and a collection of
$SU(2)$ invariant tensors (called intertwiners) $\{\iota_{n}\}$
associated to nodes $n \in \gamma$ (see left diagram on Figure
\ref{spinn}).

\section{Dynamics and spin foams}

In this section we introduce the regularization of the generalized
projection operator $P$ and state the results that are studied in detail in 
\cite{kayo0}. 

We start with the formal expression
\begin{equation} P =``\prod_{x \in \ \Sigma} \delta(\So
(x))"=\int D[N] \ {\rm exp}(i\int \limits_{\Sigma} {\rm Tr}[ N
\hat {F}(A)])\label{ppp}\end{equation}

For the moment we assume the genus of $\Sigma$ to be greater or equal
than one (the sphere case is trivial). In this case, the Riemann surface admits a global
(fiducial) flat metric $q^0_{ab}$. We introduce a square cellular
decomposition of $\Sigma$ denoted $\Sigma_{\epsilon}$ and we
require each square to have area $\epsilon^2$ with respect to
$q^0_{ab}$. We define the set of continuous finite graphs in
$\Sigma$ by $\G$. The completion in the AL norm of the set of
cylindrical functions based on $\Ge$, defined by the set of continuous
graphs contained in the 1-skeleton of the square cellular
decomposition $\Se$, defines a subspace $\He$ of the kinematical
Hilbert space $\Hk$. In the limit $\epsilon\rightarrow 0$ we have
$\He\rightarrow \Hk$.

Based on the fiducial square cellular decomposition $\Se$ we can
now define $P$ by introducing a regularization of the right hand
side of (\ref{ppp}). The integral in the exponential can be
written as
\[\int\limits_{\Sigma} {\rm Tr}[ N {F}(A)]=
\lim_{\epsilon\rightarrow 0}\ \sum_{p^i} \epsilon^2 {\rm
Tr}[N_{p^i} F_{p^i}],\] where $p^i$ labels the $i^{th}$ plaquette
and $N_{p^i}$ and $F_{pi}$ the value of $F(A)$ and $N$ at some
interior point of the plaquette respectively. The basic observation
is that the holonomy $U_{p^i}\in SU(2)$ around the plaquette $p^i$
can be written as
\[U_{p^i}[A]=\mathbbm{1}+ \epsilon^2 F_{pi}(A)+{ O}(\epsilon^2)\]
which implies
\[ \int\limits_{\Sigma} {\rm Tr}[ N {F}(A)]=\lim_{\epsilon\rightarrow 0}\ \sum_{p^i} {\rm Tr}[N_{p^i}U_{p^i}[A]].\]
Notice that the explicit dependence on the regulator $\epsilon$
has dropped out of the sum on the right hand side, a sign that we
should be able to remove the regulator upon quantization. With all
this we can define the generalized projection operator as
\begin{equation}\label{P3}\hat{P}:= \ \lim_{\epsilon\rightarrow 0} \
\ \prod_{p^i} \ \int \ dN_{p^i} \ {\rm exp}(i {\rm Tr}[ N_{p^i}
\hat {U}_{p^i}])= \lim_{\epsilon\rightarrow 0} \  \ \prod_{p^i} \
\widehat{\delta({U}_{p^i})},\end{equation} where the last equality
follows from direct integration over $N_{p^i}$ at the classical
level and $\delta(U)$ is the $SU(2)$ $\delta$-distribution 
defined on ${ L}^2(SU(2))$. We promote
$\delta(U)$ to an operator by using its expansion in $SU(2)$
irreducible representations, namely ${\delta({U_{p^i}})}=\sum_j (2j+1) \ {\chi}_j(U_{p^i})$;
${\chi}_j(U)$ is the character of the $j$-representation matrix of
$U_{p^i}\in SU(2)$. In contrast to the formal example in 4d, the previous expansion
has a precise meaning in the quantum theory
as each term in the sum can be promoted to a well defined self-adjoint operator in
$\Hk$: the Wilson loop operator in the
$j$-representation. 

Now we state the results that are proved in detail in \cite{kayo0}.

\noindent {\bf 1 -} The physical inner product $<s,s^{\prime}>_{phys}:=<sP,s^{\prime}>$ admits a {\em spin foam} representation, i.e., 
it can be expressed as a sum over {\em spin network} `histories' interpolating $s$ and $s^{\prime}$
respectively \footnote{This is obtain by the insertion of the resolution of the identity in $\Hk$ 
\[\mathbbm{1}=\sum \limits_{\gamma, \{j\}} |\gamma,\{j\}><\gamma,\{j\}|\] (where the sum is over all continuous {\em spin network states}) 
between $\delta$-distributions in (\ref{P3})
in analogy with Feynman's original derivation of the path integral representation of dynamics in QM.}. The {\em spin foams} correspond to continuous 2-complexes defined independently 
from any background structure in the limit $\epsilon \rightarrow 0$.

\noindent {\bf 2 -} The regularization (\ref{P3}) is well defined.
Lattice definitions of 3-dimensional gravity such as the 
Ponzano-Regge model are plagued with divergences. These divergences 
do not appear here however they can be traced to the presence of redundant 
$\delta$-distributions in (\ref{P3}) (see \cite{frei9} for a treatment in the 
lattice context).

\noindent {\bf 3 -} The physical Hilbert space $\Hp$ defined by the 
equivalence classes of states in $\Hk$ up to physically null states
(i.e. states $\psi$ for which $<\psi,\psi>_{phys}=0$) is isomorphic to the
Hilbert space obtained in other methods\cite{carlip}. In addition, 
the spin foam representation allows to explicitly show how {\em spin networks} 
in the same homotopy class are physically equivalent and to easily prove skein relations.

\noindent {\bf 4 -} The {\em spin foam} representation provides a 
natural basis for $\Hp$ for $\Sigma$ of arbitrary genus. These states are 
eigenstates of a complete set of commuting quantum geometry operators (provided by the formalism).
This completes the quantization of 2+1 gravity in the spin foam approach 
and establishes and establishes a clear-cut connection with the canonical picture.

\noindent {\bf 5 -} Similarly point particles can be systematically coupled to gravity\cite{kayo}. 

%\section{Discussion}

%Using the ideas presented here we can establish a clearcut
%connection between the canonical formulation of loop quantum
%gravity in three dimensions and previous covariant path integral
%definitions of the quantum theory. The Ponzano-Regge model
%amplitudes are recovered from the Hamiltonian theory and its `{\em
%continuum limit}' in the sense of Zapata \cite{za1} is built in
%from the starting point. Divergences that plage the standard
%definition of spin foam models are not present and the formalism
%provides a clear understanding of their origin which is
%complementary to the covariant analysis provided in \cite{frei9}.
%It provides an explicit realization of Reisenberger and Rovelli
%proposal for resolving dynamics in loop quantum gravity.

%We would like to emphasize the fact that we are getting fully
%background independent description of spin foam in this case. The
%generalized projection operator $P$, providing spin foam
%amplitudes, is well defined independently of any background
%structure and directly from the Hamiltonian picture. The physical
%Hilbert space is isomorphic to the .....of flat connections. Our
%result is fully consistent with the standard formulation based on
%the quantization of the reduced phase space.
We hope that the spin foam perspective, fully realized
here in 2+1 gravity, can bring new breath to the problem of
dynamics in 3+1 gravity. This is an issue that will be studied further.

%%%%%%%%%%%%%%%%%%%%%%%%%%%%%%%%%%%%%%%%%%%%%%%%%%%%%%%%%%%%%
%                                                           %
% You may repeat \section{SECTION N-th HEADING TYPE HERE}   %
%                                                           %
% Do start a subsection or sub-subsection, do this:-        %
%                                                           %
%   \subsection{SUBSECTION HEADING TYPE HERE}               %
%                                                           %
%   \subsubsection{SUBSUBSECTION HEADING TYPE HERE}         %
%                                                           %
% instead of the above                                      %
%                                                           %
%%%%%%%%%%%%%%%%%%%%%%%%%%%%%%%%%%%%%%%%%%%%%%%%%%%%%%%%%%%%%

\section*{Acknowledgments}

The authors would like to thank Lee Smolin for the kind invitation to the QST3.
This work has been supported in part by NSF Grants PHY-0090091 
and INT-0307569 and the Eberly Research Funds of Penn State.

%\bibliography{ref}

\begin{thebibliography}{0}

\bibitem{a18}
Alejandro Perez.
\newblock Spin foam models for quantum gravity.
\newblock {\em Class. Quant. Grav.}, 20:R43, 2003.

\bibitem{c2}
C.~Rovelli.
\newblock The projector on physical states in loop quantum gravity.
\newblock {\em Phys.Rev. D}, 59:104015, 1999.

\bibitem{jac}
L.~Smolin T.~Jacobson.
\newblock Nonperturbative quantum geometries.
\newblock {\em Nucl. Phys.}, B299:295, 1988.

\bibitem{c8}
C.~Rovelli L.~Smolin.
\newblock Loop space representation of quantum general relativity.
\newblock {\em Nucl. Phys. B}, 331:80, 1990.

\bibitem{ash3}
A.~Ashtekar and J.~Lewandowski.
\newblock Projective techniques and functional integration.
\newblock {\em J. Math. Phys.}, 36:2170, 1995.

\bibitem{ash4}
A.~Ashtekar and J.~Lewandowski.
\newblock Representation theory of analytic holonomy $c^*$ algebras.
\newblock {\em in ``Knots and quantum gravity'', J. Baez (ed), Oxford
  University Press, Oxford 1994}.

\bibitem{reis8}
Michael~P. Reisenberger.
\newblock World sheet formulations of gauge theories and gravity.
\newblock {\em gr-qc/941235}, 1994.

\bibitem{c4}
C.~Rovelli and L.~Smolin.
\newblock Spin networks and quantum gravity.
\newblock {\em Phys. Rev. D}, 53:5743, 1995.

\bibitem{baez10}
J.~Baez.
\newblock Spin network states in gauge theory.
\newblock {\em Adv.Math.}, 117:253--272, 1996.

\bibitem{kayo0}
K.~Noui A.~Perez.
\newblock Three dimensional loop quantum gravity: physical scalar product and
  spin foam models.
\newblock {gr-qc/0402110}.

\bibitem{frei9}
Laurent Freidel and David Louapre.
\newblock Diffeomorphisms and spin foam models.
\newblock {\em Nucl. Phys.}, B662:279--298, 2003.


\bibitem{carlip}
S.~Carlip.
\newblock Quantum gravity in 2+1 dimensions.
\newblock Cambridge, UK: Univ. Pr. (1998) 276 p

\bibitem{kayo}
K.~Noui A.~Perez.
\newblock Three dimensional loop quantum gravity: coupling with point particles.
\newblock {gr-qc/0402111}.

\end{thebibliography}
%\bibliographystyle{unsrt}
%\end{document}

\end{document}